\documentclass[10pt,a4paper,notitlepage]{article}
\usepackage[utf8]{inputenc}
\usepackage[english]{babel}
\usepackage{amsmath}
\usepackage{amsfonts}
\usepackage{amssymb}
\usepackage{mathbbol}
\usepackage{graphicx}
\usepackage{float}
\usepackage{wrapfig}
\usepackage{caption}
\usepackage{subcaption}
\usepackage{enumerate} 
\usepackage[hidelinks]{hyperref}
\usepackage{cite}
\usepackage{xcolor}
\usepackage{setspace}
\usepackage{enumerate}

\usepackage{titling}
\usepackage{lipsum}
\usepackage[left=1in, right=1in, top=1in, bottom=1in]{geometry}
\usepackage{authblk}

\title{Persistent Threshold Dynamics
 with Recovery in Complex Networks}
\author[1]{{\small Nanxin Wei}}
\affil[1]{{\scriptsize \textit{Department of Mathematics , Imperial College London}}}
\author[2]{{\small Bo Fan}}
\affil[2]{{\scriptsize \textit{Department of Statistics, University of Oxford}}}
\date{\today}
\begin{document}

\maketitle
\thispagestyle{empty}

\begin{abstract}
Threshold rules of spreading in binary-state networks lead to cascades. We study persistent cascade-recovery dynamics on 
\textit{quasi-robust} networks, i.e., networks which are robust against small trigger but may collapse for larger one.
It is observed that depending on the relative rate of triggering and recovery, the network falls into one of the two dynamical 
phases: collapsing or active phase. We devise an analytical framework which characterizes not only the critical behavior but also 
the temporal evolution of network activity in both phases. Agent-based simulation results show good agreement with the analytical 
calculations, \textbf{indicating strong predicative power of our method for persistent cascade dynamics in complex networks.} 
\end{abstract}

\newpage
\section{Introduction} 
\paragraph*{}
Spreading processes has been a center of focus in the studies of dynamics on networks in recent years \cite{newman2010networks,porter2016dynamical}. One type of particular interest is called \emph{threshold model} (or \emph{threshold contagion}), in which spreading is activated or facilitated only if the local drive surpasses certain threshold. Threshold models are intimately related to cascading phenomena prevalent in both natural and artificial complex systems, where small trigger induces consecutive responses that spread through the system. Examples of cascading can be found in blackouts of power grids \cite{motter2002cascade,carreras2004evidence}, defaults in financial networks \cite{roukny2013default,hurd2016double}, firing in neural systems \cite{stein1967some,koch1998methods}, and information spreading in social networks \cite{centola2007cascade,gao2016general}. The introduction of threshold model in social theories dates back to Schelling and Granovetter \cite{schelling1969models,schelling1973hockey,granovetter1978threshold}, while its recent gained popularity in network science largely owes to the seminal work of Watts \cite{watts2002simple}, where he formulated the "Watts threshold model" (WTM) and derived the \emph{global cascade criterion}, i.e., the combination of network topology and threshold condition that admits global cascade for small initial trigger (infinitesimal in the infinite network limit). Many following studies has been attempted for variations of the WTM, mostly focusing on global cascades in the context of network resilience and outburst of information, in parallel with the studies of percolation on networks \cite{newman2010networks}.
An important advancement in the theoretical analysis of WTM was made by Gleeson et al. \cite{gleeson2007analytical,gleeson2007seed,gleeson2017message}. Using message-passing techniques they established the effective dynamics during threshold driven cascades under initial trigger (referred to as \emph{seeds} in the above literatures) and improved Watts' global cascade criterion. Gleeson's method was developed for locally tree-like random networks, and has since been generalized to more complicated network topologies \cite{gleeson2008cascades,hackett2011cascades,hackett2013cascades,hurd2013watts}.

\paragraph*{}
Some real-world networks, while being less susceptible to global cascade for small initial trigger, may be subjected to persistently triggered cascades and some recovery mechanism. Typically, cascades happen in a much shorter timescale than the intermittency of trigger or recovery. Such systems follow the dynamical patterns as: \emph{trigger$\rightarrow$cascades($\rightarrow$recovery)$\rightarrow$trigger$\rightarrow\cdots$}\footnote{Or \emph{$\cdots\rightarrow$trigger/recovery$\rightarrow$cascades$\rightarrow\cdots$}}, which are observed in a variety of social \cite{singh2013threshold,karsai2014complex,adamic2015diffusion,alvarez2015sentiment}, technical \cite{schlapfer2009modeling,buldyrev2010catastrophic,johansson2011vulnerability,quattrociocchi2014self}, financial \cite{majdandzic2014spontaneous,podobnik2014network,podobnik2014systemic} and biological networks \cite{mattern2004imbalance,vural2014aging}. The original WTM with one-off trigger can be viewed as the first step in the persistent picture: \emph{trigger$\rightarrow$cascades}. Therefore, to characterize \emph{persistent threshold models} is of significant importance not only as a generalization of WTM, but also in understanding and predicting a large class of complex systems. However, due to the non-monotonicity and separation of timescales of the dynamics, not much theoretical progress has been made. \cite{majdandzic2014spontaneous} devised a threshold model with persistent internal failure (trigger), external failure (cascades) and deterministic recovery, and showed that it incorporated richer behaviour than in the one-off trigger scenario \cite{podobnik2014network,podobnik2014systemic,podobnik2015predicting}. Recently \cite{valdez2016failure,bottcher2016impact,bottcher2017failure,bottcher2017critical} considered models with fixed thresholds and stochastic recovery in continuous time dynamics. 

\paragraph*{}
Here, we investigate a minimal persistent threshold model with stochastic trigger and recovery on a general class of random networks (which we call \emph{quasi-robust} networks, see Section 3 for details), without the \emph{ad hoc} simplifications in previous studies. Dependent on the relative intensity of recovery over trigger as control parameter, two phases: \emph{active} and \emph{collapsing} phase are identified (see Figure 1) and analyzed in a unified theoretical framework. The theory provides a sufficient bound condition for the network to sit in the collapsing phase and the time of collapse when it does, as well as predicting the evolution of the network with good accuracy in both phases. The model is very generic and the theoretical approach can be easily extended to more complicated set-ups. Our contribution is an important step in exploring networked systems which slowly evolve to the point of critical collapse (or outburst in terms of spreading).

\paragraph*{}
The rest of the article will be organized as follows. In Section 2 we define the model and describe the general behaviours of the networks of interest. In Section 3 we build on Gleeson's method \cite{gleeson2007analytical,gleeson2007seed,gleeson2017message} and perform a \emph{trigger size analysis} of WTM, identifying the class of quasi-robust random networks that is expected to show the observed phenomena in the persistent dynamical picture. Using the results from one-off trigger size analysis, in Section 4 we construct an effective process that mimic the full dynamics of the model, approximating the temporal evolution of the network activity with a rate equation of \emph{effective trigger size} and accompanied threshold cascades. In Section 5 we compare simulation results with theoretical calculations for some key observables and discuss the predictive power and limitations of our framework. Section 5 concludes with a brief summary of current results, potential applications and future work. 

\begin{figure}[h]
\centering
\begin{subfigure}{.5\textwidth}
  \centering
  \includegraphics[width=1.0\linewidth]{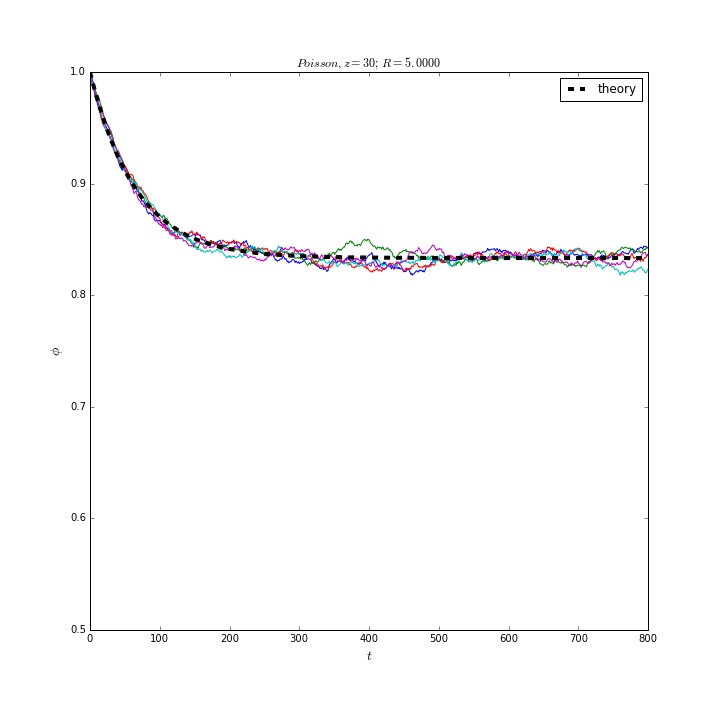}
  \caption{\scriptsize{Active phase, high recovery}}
  \label{fig:sub1}
\end{subfigure}%
\begin{subfigure}{.5\textwidth}
  \centering
  \includegraphics[width=1.0\linewidth]{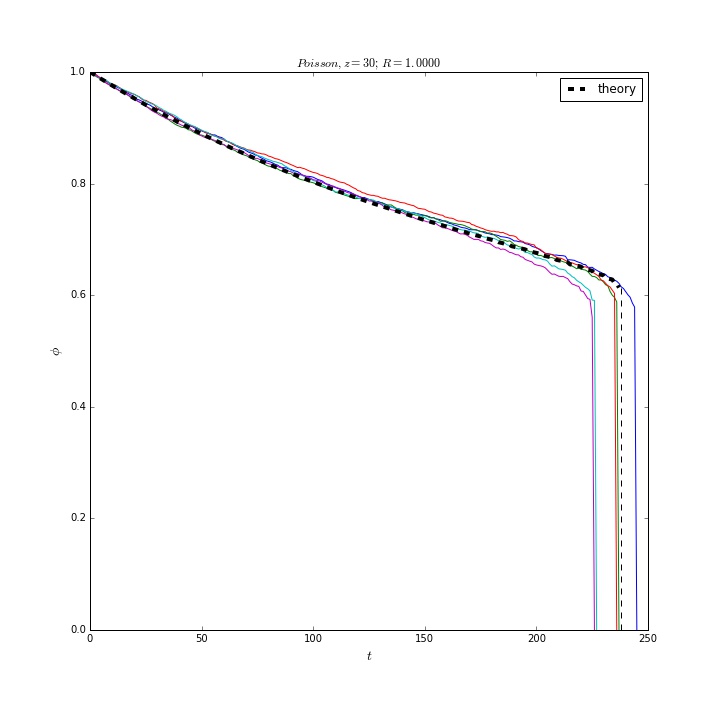}
  \caption{\scriptsize{Collapsing phase, low recovery}}
  \label{fig:sub2}
\end{subfigure}
\caption{Activity evolution for two phases}
\medskip 
\begin{minipage}{0.8\textwidth} 
{\footnotesize The temporal evolution of network activity $\phi$ for threshold dynamics with recovery on Poisson random networks with mean degree $z=30$, uniform threshold condition $\theta=0.5$, trigger rate $\lambda_{0}=0.0025$ and recovery rate $\lambda_{1} = R\cdot\lambda_{0}$ ; (a) $R=5.0000$ (b) $R=1.0000$. Starting from fully active initial configuration $\phi = 1$, five realizations are shown as colourful thin lines in both (a)\&(b), while black thick dash lines are theoretical calculations given by the following sections.}
\end{minipage}
\label{fig:test}
\end{figure}

\section{Model}
\paragraph*{}
Consider discrete-time threshold dynamics with recovery on random networks (configuration model conditioned on simple graphs \cite{newman2010networks}). To be aligned with some related studies, we borrow the terminologies from technical networks and refer to the binary state of each node as \emph{active} or \emph{failed}. An active node can fail spontaneously, failure spreads according to (fractional) threshold rule, and a failed node can recover to be active (spontaneously). Precise set-up of the model is given as follows. \footnote{Treating spontaneous failure and recovery equally (while cascading failure still happens in a much shorter time-scale) we would have a different model set-up, which is totally consistent with the current analytical framework. Such set-up would also be naturally consistent with the continuous time version of the model, where $\lambda_{0}$ and $\lambda_{1}$ are truly rates of events.}

\begin{enumerate}
\item 
Random networks of given size $N$ and degree distribution $\{p_{k}\}$;
\item Two dynamical parameters: \\
Spontaneous failure (trigger) rate $\lambda_{0}\ll 1 $ and recovery rate $\lambda_{1}\ll 1 $;
\item Uniform fractional threshold $\theta= 0.5$ for failure spreading; 
\item Dynamics: \\
Network starts from fully active initial configuration, going through procedures (i)-(iv) in each \emph{time-step}. Updates of the network are referred to as \emph{micro-steps}. 
\begin{enumerate}[(i)]
\item Each active node is independently subject to spontaneous failure with probability $\lambda_{0}$; 
\item If the failure fraction of the neighbours of an active node exceeds threshold $\theta$, the node is damaged in the next micro-step. 
Such cascading failure continues until all active nodes fulfil the sustainable condition, i.e., having no more than $\theta$ fraction of failed neighbours;
\item Each failed node is independently subject to recovery with probability $\lambda_{1}$ (synchronous update); 
If some recovered nodes don't fulfil the sustainable condition, they shall fail again in the next micro-step, which may yet be followed with cascades until no updates (unsustainable recovery);\footnote{Asynchronous update could incur further complications. For the final network configuration to be independent of the order of updates, the survivability check should be set at the beginning of each time-step. But as we shall argue later, the difference between update schemes is negligible under approximation.}
\item Start from (i) and iterate. 
\end{enumerate}
\end{enumerate} 

\paragraph*{} 
Define $R \triangleq \lambda_{1} / \lambda_{0}$ (relative intensity of recovery over trigger), $\phi$ as the network activity (fraction of active nodes in the network), and $\rho = 1-\phi$ the corresponding failure fraction. For a large class of networks, the temporal evolution of $\phi$ generally exhibits one of the two types of behaviour dependent on chosen $R$: decreasing to an intermediate value and fluctuating around it for large $R$, or gradually dropping before suddenly collapsing to much lower value for small $R$. We refer to them as the \emph{active phase} and the \emph{collapsing phase} respectively, as shown in Figure 1.

\section{Trigger size analysis of Watts threshold model}
\paragraph*{}
As the persistent trigger threshold dynamics with recovery defined in Section 2 is hard to tackle directly, we look into the one-off trigger scenario - the Watts threshold model \cite{watts2002simple} first, in the attempt that it could serve as a building block for the analysis of the full dynamics. By extending Gleeson's method \cite{gleeson2007analytical,gleeson2007seed,gleeson2008cascades} to a complete trigger size analysis, we show that even for uniform threshold condition, networks may present richer behaviour than that depicted in the global cascade criterion.  

\paragraph*{}
Consider the same random network ensemble as described in Section 2, each of $N$ nodes, degree distribution $p_{k}$ and mean degree $\langle k \rangle=\sum_{k=1}^{\infty}kp_{k}=z \ll N$. Assume the network is locally tree-like, which is naturally satisfied for most weakly connected large random networks, including all of the situations discussed in this paper. From the fully active configuration, a fraction $\eta$ of nodes are chosen to be initial failures ($\eta$ as \emph{trigger size}). The probability of an arbitrary node being failed after all cascades is the (expected) final failure fraction of the network $\rho_{\bullet}$. In order to estimate $\rho_{\bullet}$, approximate the network as a rooted tree branching from the arbitrarily chosen probe node, which is connected to $k$ children with probability $p_{k}$. Each of children nodes is in turn connected to $(k_{i}-1)$ children with probability $\tilde{p}_{k}=kp_{k}/z$ (the extended degree distribution), as are all the nodes in ensuing levels. Label the levels of the resulting tree structure from bottom up as $n=0,1,2...$ . In the infinite network limit $N\rightarrow\infty$, the level number of the root $n_{max}\rightarrow\infty$. 

\paragraph*{}
Since the final network configuration is independent of the order of cascading failure updates, let cascading spread from bottom to top of the tree. Denote $\nu_{n}$ as the probability of a node being failed on condition that its parent is not initially failed.
Since initial failures are independent events, we have iterative relation:
\begin{equation}\label{eq1}
\nu_{n+1}=\eta+(1-\eta)\sum_{k=1}^{\infty}\dfrac{k}{z}p_{k}\sum_{m=0}^{k-1}{k-1 \choose m}\nu_{n}^{m}(1-\nu_{n})^{k-1-m}F(m,k) \qquad n=0,1,2,\ldots 
\end{equation}
Here $F(m,k)$ is the spreading (response) function of a node with degree $k$ and $m$ failed children. For fractional threshold rule with threshold $\theta$, 
\begin{equation}\label{eq2}
    F(m,k)= 
\begin{cases}
    0,& \text{if } m\leqslant \theta k\\
    1,& \text{if } m> \theta k
\end{cases}
\end{equation}
Take $n\rightarrow\infty$, Eq.(\ref{eq1}) leads to a self-consistent equation about $\nu\triangleq \nu_{\infty}$:
\begin{equation}\label{eq3}
\nu=\eta+(1-\eta)g(\nu)
\end{equation}
where 
\begin{equation}\label{eq4}
g(\nu)=\sum_{k=1}^{\infty}\dfrac{k}{z}p_{k}\sum_{m=0}^{k-1}{k-1 \choose m}\nu^{m}(1-\nu)^{k-1-m}F(m,k)
\end{equation}

\paragraph*{}
Similarly, the probability of the root node being failed is given by
\begin{equation}\label{eq5}
\rho_{\bullet}=\eta+(1-\eta)h(\nu)
\end{equation}
where 
\begin{equation}\label{eq6}
h(\nu)=\sum_{k=1}^{\infty}p_{k}\sum_{m=0}^{k}{k \choose m}\nu^{m}(1-\nu)^{k-m}F(m,k)
\end{equation}
From Eq.(\ref{eq5}) and the physically viable solution of Eq.(\ref{eq3}) $\nu(\eta)$ (to be elaborated in the next paragraph) we can establish the mapping from $\eta$ to $\rho_{\bullet}$: $\rho_{\bullet}(\eta) \doteq \rho_{\bullet}(\eta,\nu(\eta))$.

\paragraph*{}
For different network degree distributions, Eq.(\ref{eq3}) may incorporate qualitatively different dynamical behaviours. Note that from definition $\eta\in[0,1]$, $\nu\in[0,1]$, $g(0)=0$, $g(1)=1$, $g(\nu)\in(0,1)$ for $\nu \in(0,1)$, and the initial condition of iteration $\nu_{0} = \eta$, the smallest solution of Eq.(\ref{eq3}) given $\eta$ would be the physical one.\footnote{The first fixed point of the iterating equation Eq.(\ref{eq1}) to be reached; And if $\nu(\eta)<1$, it is always a stable fixed point, see \cite{gleeson2007analytical}.} The global cascade criterion studied in \cite{watts2002simple,gleeson2007analytical,gleeson2007seed,gleeson2008cascades} corresponds to the scenario where infinitesimal initial trigger would cause finite fraction of final failure, in which case $\nu \equiv 1$ for any $\eta\in(0,1]$. From the view of network resilience, such networks could be considered as \emph{fragile}. Outside the fragile regime, Eq.(\ref{eq3}) has solution(s) $\nu(\eta) \in (0,1)$ for $\eta \in (0,1)$. If there's unique solution $\nu(\eta)$ as $\eta$ grows from 0 to 1, since $g(\bullet)$ is a continuous function, $\nu(\eta)$ should be continuous as well. In such networks no sudden surge of failure fraction can be observed as the trigger size increases\footnote{From Eq.(\ref{eq5})\&Eq.(\ref{eq6}) it is obvious that $\rho_{\bullet}(\eta)$ and $\nu(\eta)$ follows similar pattern.}, which we would dubbed as \emph{robust}. If for some $\eta\in(0,1)$, Eq.(\ref{eq3}) has more than one solution of $\nu(\eta) \in (0,1)$, taking the the smallest one as physical solution leads to a discontinuous jump in the $\eta$-$\nu$ relation. This type of \emph{quasi-robust} networks that entails a surge of failure fraction (hence collapse in activity) when the initial trigger reaches certain size is the focus of this work\footnote{The characterization of network robustness based on its structure remains an open problem to be further explored.}, which we expect to present similar properties in the persistent dynamics. To illustrate the discussions above, Figure 2 depicts two $\rho_{\bullet}(\eta)$ curves for robust and quasi-robust networks respectively.
\begin{figure}[h]
\centering
\includegraphics[width=.5\linewidth]{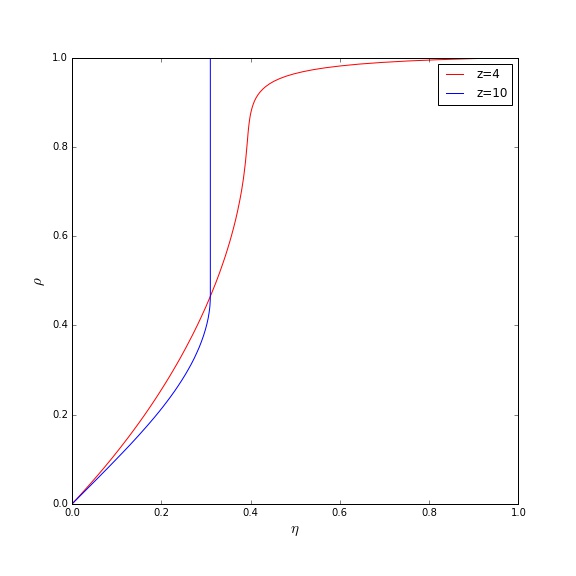}
\caption{Robust and quasi-robust random networks \\}
\label{fig:eta-rho}
\medskip 
\begin{minipage}{0.8\textwidth} 
{\footnotesize Expected final failure fraction $\rho_{\bullet}$ in Watts threshold model over varying initial trigger size $\eta$, on Poisson random networks with mean degree $z=4$ (red) and $z=10$ (blue), uniform fractional threshold $\theta=0.5$}
\end{minipage}
\end{figure}

\paragraph*{}
In the quasi-robust networks of interest (such as $z=10$), we can also determine the point of critical surge in failure fraction, or critical collapse in vitality. Critical surge corresponds to $\delta\rho_{\bullet} \gg \delta\eta$, which happens if and only if $\delta\nu\gg\delta\eta$. Take variation of Eq.(\ref{eq3}) at both sides, we have
\begin{align}
&\delta\nu = \delta\eta - \delta\eta g(\nu) + (1-\eta)g'(\nu)\delta\nu \notag \\
\Longrightarrow \qquad &\dfrac{\delta\eta}{\delta\nu} = \dfrac{1-(1-\eta)g'(\nu)}{1-g(\nu)} \label{eq7}
\end{align}
identify the critical point as $\dfrac{\delta\eta}{\delta\nu}\bigg\vert_{crit}\rightarrow 0$ and note that $\eta=\dfrac{\nu-g(\nu)}{1-g(\nu)}$ approaching the critical point from below,
\begin{align}
&\frac{1-\nu^{(c)}}{1-g(\nu^{(c)})}g'(\nu^{(c)})=1 \notag \\
&\eta^{(c)}=\dfrac{\nu^{(c)}-g(\nu^{(c)})}{1-g(\nu^{(c)})} \label{eq8}
\end{align}
are the critical point equations given by trigger size analysis.

\section{Effective trigger size evolution of threshold dynamics with recovery}
\paragraph*{}
In order to theoretically characterize the persistent process on quasi-robust networks, it would be tempting to replace the increasing trigger size $\eta$ in the analysis in Section 3 with the temporal evolution of some effective trigger size $\eta_{e}(t)$. Here we argue that by classifying the failed nodes in persistent dynamics into two types: the ones spontaneously failed and the ones failed by cascading (short as \emph{spon-failed} and \emph{cas-failed}), under certain assumptions the spon-failed nodes can indeed serve as effective trigger $\eta_{e}(t)$ to estimate the network activity $\bar{\phi}(t)$.

\paragraph*{}
Assume that when the network (either in active or collapsing phase) evolves in low failure regime, spon-failed nodes that fall below the sustainable condition are negligible. This means that all spon-failed nodes are recoverable. Further we assume that all cas-failed nodes at $t$ (failed by cascading over time) are still below the sustainable condition. Recall that $\lambda_{0},\lambda_{1} \ll 1$, the second assumption would thus infer that cas-failed nodes are much less likely to be recovered (or may only recover at much slower pace) compared with spon-failed nodes. The above assumptions are reasonable in all cases considered in this work, and are believed to hold for quasi-robust networks in general.

\paragraph*{}
The theoretical estimation of the persistent process is given by the following \emph{effective process}: In each time-step, (i) Each active node turns spon-failed with probability $\lambda_{0}$ and each spon-failed node recovers with probability $\lambda_{1}$ (independently); (ii) The updated spon-failed nodes serve as the \emph{effective trigger} at current time and decide the cas-failed nodes according to the threshold dynamics. In other words, it is the alternating combination of "independent spontaneous failure/recovery of active/spon-failed nodes" and "refreshing of cascading failure". It is worth stressing that the different treatment of two types of failed nodes here is not because of imposed intrinsic property of failure, as was the case in \cite{podobnik2014network}), but is a result due to their different positioning in the network hence their susceptibility to recovery. Figure 3 
provides a schematic illustration of the effective dynamics.

\begin{figure}[h]
\centering
\includegraphics[width=.6\linewidth]{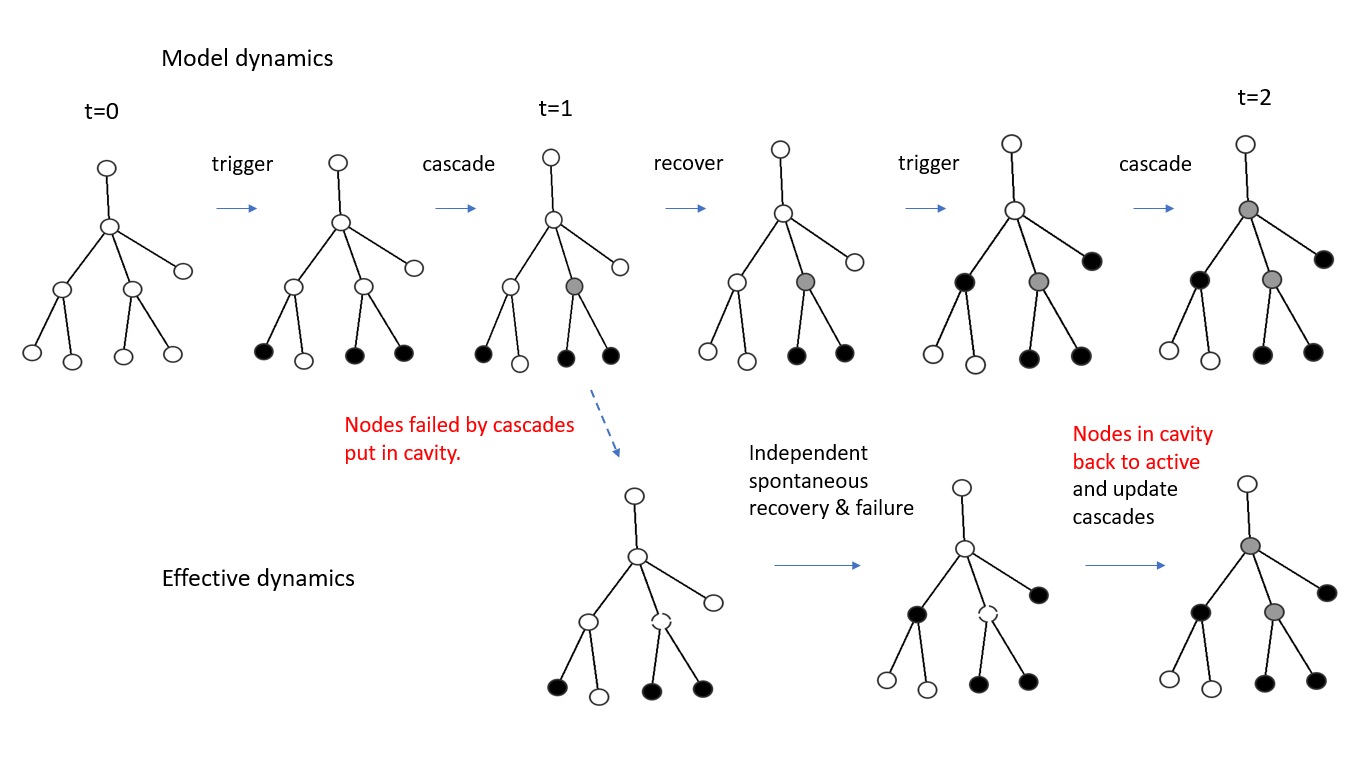}
\caption{Schematic illustration of the effective dynamics in contrast to the model dynamics\\}
\label{fig:schematic}
\medskip 
\begin{minipage}{0.8\textwidth} 
{\footnotesize Example of a possible realisation of the model dynamics (upper panel) and the corresponding effective dynamics (lower panel). 
Empty nodes with solid boundary are active nodes, black for spon-failed nodes, grey for cas-failed nodes. Empty nodes with dashed boundary 
are cas-failed nodes put in cavity, i.e., perceived as active nodes during the updates between two macroscopic time-steps.}
\end{minipage}
\end{figure}

\paragraph*{}
Using results from the trigger size analysis for expected cascading failure, the rate equation of the effective trigger size $\eta_{e}(t)$ is given by
\begin{equation}\label{eq9}
\dfrac{\mathrm{d}\eta_{e}}{\mathrm{d}t}=\lambda_{0}(1-\rho_{\bullet}(\eta_{e}))-\lambda_{1}\eta_{e}
\end{equation}
with the initial condition $\eta_{e}(0)=0$. The estimation of network activity at time $t$ (before critical collapse for networks in the collapsing phase or without constraint in the active phase) would thus be $\bar{\rho}(t) \doteq \rho_{\bullet}(\eta_{e}(t))$, where $\eta_{e}(t)$ is the solution of Eq.(\ref{eq9}).

\paragraph*{}
If the network belongs to the active phase, it is expected to enter a steady state where $\mathrm{d}\eta_{e}/\mathrm{d}t=0$. Denote $\eta_{e}^{*} \triangleq \eta_{e}(\infty)$, then
\begin{equation}\label{eq10}
\rho_{\bullet}(\eta_{e}^{*})=1-R\eta_{e}^{*}
\end{equation}
$R = \lambda_{1} / \lambda_{0}$ as defined earlier. Eq.(\ref{eq10}) shows that in our analysis the expected steady-state activity $\bar{\rho}$ only depends on the relative intensity of recovery over trigger. For given high $R$ (ensuring active phase), the solution of Eq.(\ref{eq10}) $\eta_{e}^{*}(R)$ and hence $\bar{\rho}(R) \doteq \rho_{\bullet}(\eta_{e}^{*}(R))$ can be found numerically, as demonstrated in Figure 4 the intersection of left hand side $\rho_{\bullet}(\eta)$ (blue line) and right hand side $1-R\eta$ (red dashed line).
\begin{figure}[h]
\centering
\includegraphics[width=.5\linewidth]{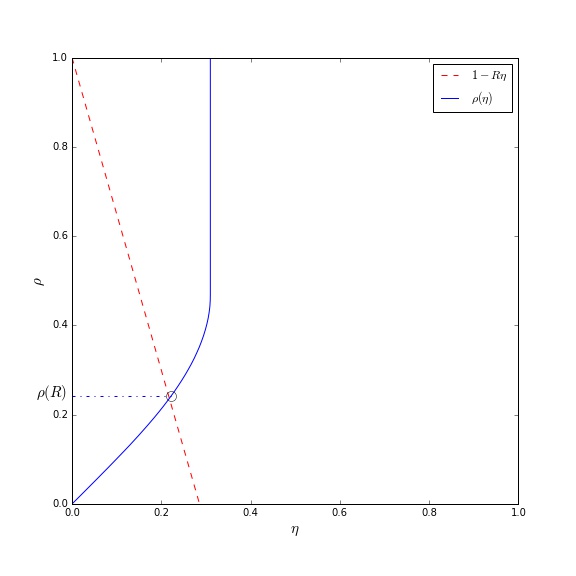}
\caption{$\rho(R)$ from effective trigger size of steady state \\}
\label{fig:R-rho}
\medskip 
\begin{minipage}{0.8\textwidth} 
{\footnotesize Dynamics on Poisson random networks with degree distribution $p_{k}=e^{-z}z^{k}/k!$, $z=10$ and fractional threshold $\theta=0.5$, the estimation of steady state death fraction $\rho$ given $\lambda$-ratio $R$, shown here for $R=3.5$. }
\end{minipage}
\end{figure}

\paragraph*{}
Figure 4 also shows that $\eta_{e}^{*}$ monotonically increases as $R$ decreases. Take the theoretical value of critical $R$ as 
\begin{equation}\label{eq11}
\bar{R}^{(c)}\triangleq [1-\rho_{\bullet}(\eta^{(c)})]/\eta^{(c)}
\end{equation}
where the critical trigger size $\eta^{(c)}$ is given by Eq.(\ref{eq8}). The network is expected to be in active phase when $R>R^{(c)}$ and in collapsing phase when $R<R^{(c)}$. For network in the collapsing phase, failure fraction would evolve in the form of $\rho_{\bullet}(\eta_{e})$ as $\eta_{e}(t)$ increases until reaching $\eta^{(c)}$. The time of collapse $t^{(c)}$ would therefore be identified as the solution of the boundary value problem of Eq.(\ref{eq9}) with end-time condition $\eta_{e}(\bar{t}^{(c)})=\eta^{(c)}$.

\section{Results}
\paragraph*{}
Calculations based on the effective process introduced in the last section are conducted to compare with agent-based simulation results. Estimations for the temporal evolution of network vitality are plotted in black thick dash lines in Figure 1, in good agreement with simulation results in both the active phase and pre-collapse stage of the collapsing phase.

\paragraph*{}   
Take a decreasing set of large $R$ values and record the averaged steady-state activity for each $R$. Results are shown by the red dots in Figure 5, with detailed parameters in figure comments. Blue lines are plotted through the theoretical estimations $\bar{\rho}(R)$ of the same set of $R$ values. Theory and simulations agree with convincing accuracy for networks safely in the active phase ($R \gg \bar{R}^{(c)}$), regardless of network structure. As $R$ approaches the theoretical critical point $\bar{R}^{(c)}$ from above, deviations for the steady-state activity kick in. Better fit for Poisson random network with mean degree $z=30$. 
Nevertheless, in all cases the actual critical point $R^{(c)}$ found in simulations is larger than its theoretical value $\bar{R}^{(c)}$. In other words, theoretical estimation $\bar{R}^{(c)}$ provides a lower bound, below which the network would safely belong to the collapsing phase. This can be explained by further scrutiny over the effective process, and comparison with the actual process. As the theoretical failure size $\bar{\rho}(t)$ (for any given $R$) is derived by the effective trigger size, i.e., the fraction of spon-failed nodes, the \emph{recoverable cas-failed} nodes (nodes killed by cascades earlier, but at current time-step fulfil the sustainable condition due to neighbour recovery over time) are considered to be recovered with \emph{certainty} in the effective process, but not necessarily so in the actual process. For lower $R$, failure size would increase such that this fraction of nodes becomes no longer negligible. Therefore, it is to be expected that the theoretical failure size is an \emph{underestimation}, hence in simulations the networks are prone to collapse for some $R > \bar{R}^{(c)}$.

\begin{figure}[h]
    \centering
    \begin{subfigure}[b]{0.30\textwidth}
        \includegraphics[width=\textwidth]{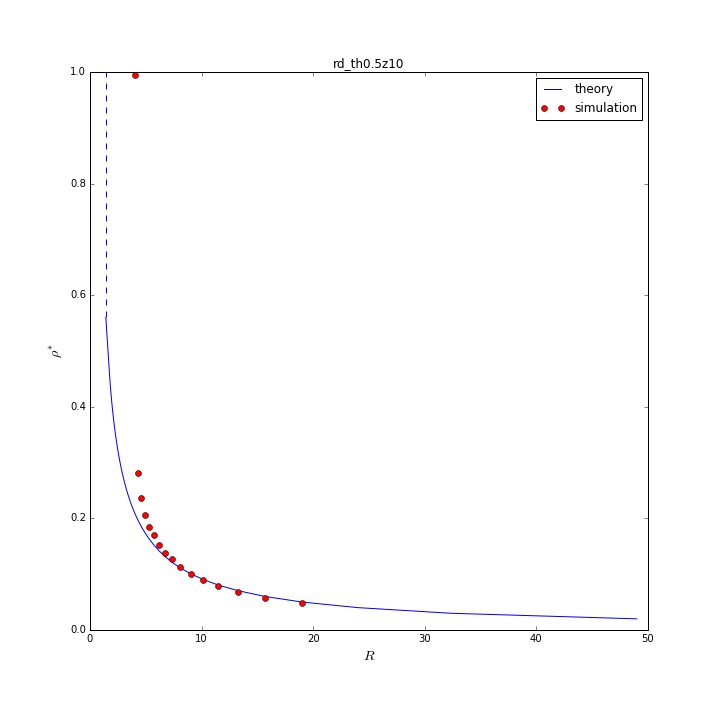}
        \caption{}
        \label{fig:rd_z10}
    \end{subfigure}
    ~ 
    \begin{subfigure}[b]{0.30\textwidth}
        \includegraphics[width=\textwidth]{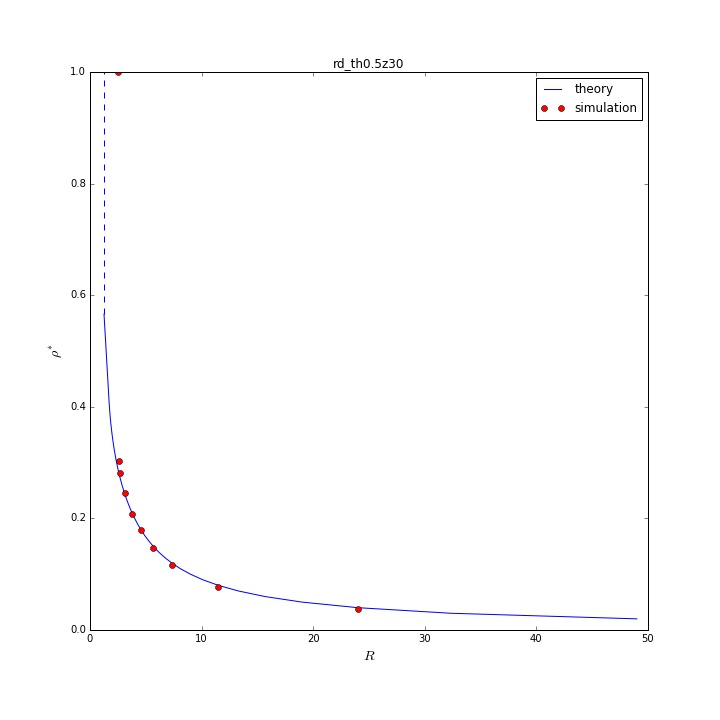}
        \caption{}
        \label{fig:rd_z30}
    \end{subfigure}
    \begin{subfigure}[b]{0.30\textwidth}
        \includegraphics[width=\textwidth]{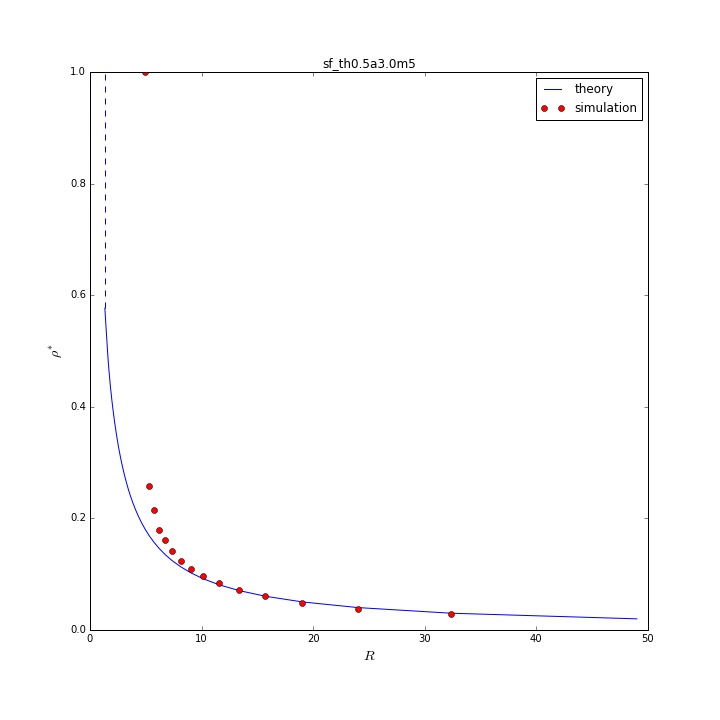}
        \caption{}
        \label{fig:sf_a3m5}
    \end{subfigure}
    \caption{$\rho(R)$ in the active phase}
    \label{fig:compareRho}
	\begin{minipage}{0.8\textwidth} 
	{\footnotesize (a) \& (b) Poisson random networks with $z=10$ and $z=30$ respectively; (c) Scale-free random networks with  $p_{k}\propto k^{-\gamma}$, $\gamma=3.0$, $k_{min}=5$, $\langle k\rangle \sim 8.7$. In all cases we fix $\theta=0.5$ and $\lambda_{0}=0.0025$.}
	\end{minipage}
\end{figure}

\paragraph*{}   
For networks in the collapsing phase ($R < \bar{R}^{(c)}$), we can also calculate the time of collapse $\bar{t}^{(c)}$, and compare with $t^{(c)}$ extracted from simulations, as shown in Figure 6. Blue lines plotted from theoretical estimations, while red dots with error bars are average and standard deviation of $t^{(c)}$ from simulations. In all cases theory and simulations are in reasonable alignment, with $\bar{t}^{(c)}(R)$ being an expected upper bound of $t^{(c)}(R)$. Following the same argument in the last paragraph, since the estimated failure size being underestimation, the actual time of collapse comes earlier than the theoretically predicted time.
 
\begin{figure}[h]
    \centering
    \begin{subfigure}[b]{0.30\textwidth}
        \includegraphics[width=\textwidth]{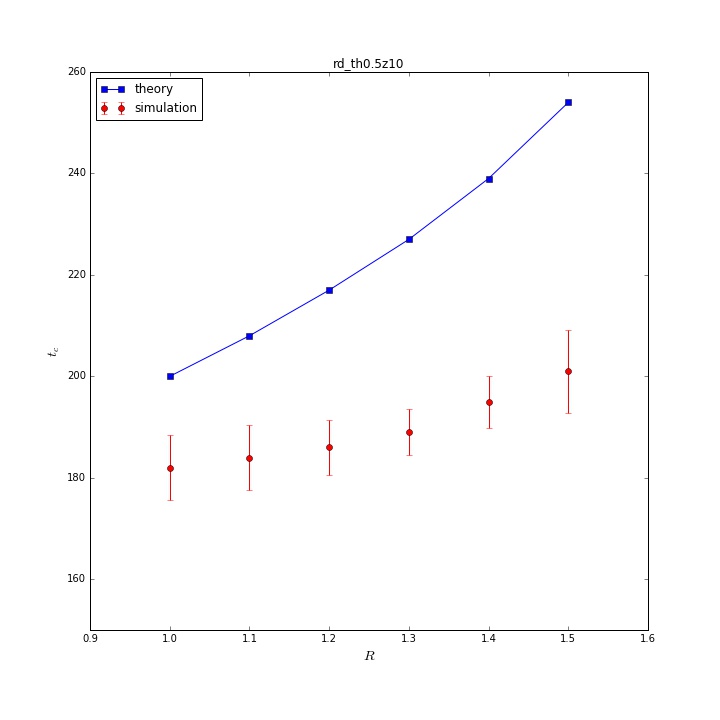}
        \caption{}
        \label{fig:rd_z10}
    \end{subfigure}
    ~ 
    \begin{subfigure}[b]{0.30\textwidth}
        \includegraphics[width=\textwidth]{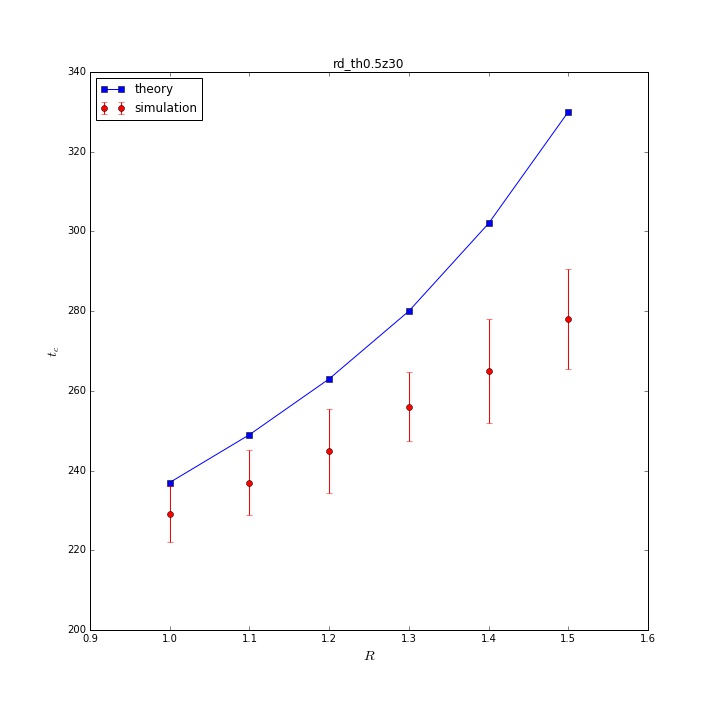}
        \caption{}
        \label{fig:rd_z30}
    \end{subfigure}
    \begin{subfigure}[b]{0.30\textwidth}
        \includegraphics[width=\textwidth]{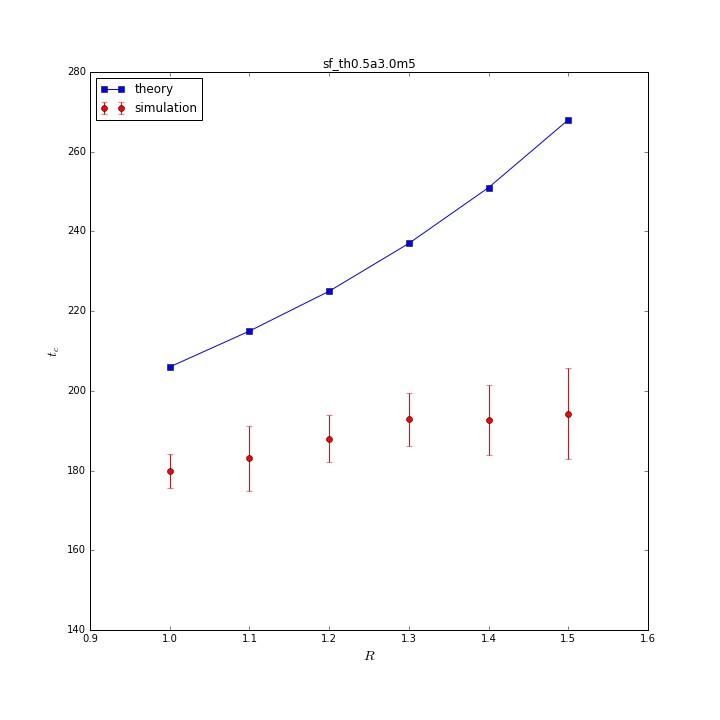}
        \caption{}
        \label{fig:sf_a3m5}
    \end{subfigure}
    \caption{$t^{(c)}(R)$ in the collapsing phase}
    \label{fig:compareRho}
	\begin{minipage}{0.8\textwidth} 
	{\footnotesize (a) \& (b) Poisson random networks with $z=10$ and $z=30$ respectively; (c) Scale-free random networks with  $p_{k}\propto k^{-\gamma}$, $\gamma=3.0$, $k_{min}=5$, $\langle k\rangle \sim 8.7$. In all cases we fix $\theta=0.5$ and $\lambda_{0}=0.0025$.}
	\end{minipage}
\end{figure}


\section{Conclusions}
\paragraph*{}
In this paper we developed a method to calculate the average activity value of complex networks under cascade-recovery dynamics if the ratio of recovery/death rates is high enough to keep the network in active phase, and calculate the collapsing time when the drastic drop of activity of the network happened in the low ratio case. The simulation results of uncorrelated networks are quantitatively consistent with our theoretical predication, where moderate discrepancies near critical point were witnessed and could be interpreted by the optimistic estimation of recoverable nodes killed by threshold cascading rule in previous steps. The critical ratio $R^{(c)}$ predicted by our method, therefore, can be used as a lower bound in practice, and the predicted collapsing time $t^{(c)}$can serve as a upper bound of the longevity of the system, respectively.

\paragraph*{}
From the perspective of application, our model can be potentially applied to predict the collapsing time of complex systems with interdependency, or the condition for it to remain in active phase with high activity (i.e. high portion of functional components) in terms of the ratio of functional/dysfunctional rates on each component. An example is the stock market in which the performance of the stock of each company depends on its upstream-downstream partners, while the 0 and 1 states of each company correspond to debt deficit status and debt solvent status respectively \cite{majdandzic2014spontaneous,podobnik2014network,podobnik2014systemic}. Our method then can be used to predict whether the collapse of the entire stock market could be triggered by the chain reaction of debt deficits of listed companies based on enterprise financing efficiency and disbursement rate. Similarly, with our method we could also try to predict the lifespan of some animal \cite{vural2014aging}, the collapsing time and condition of honey bee colonies \cite{perry2015rapid}, the failure time of power grids \cite{wang2009cascade}, the behaviour of progression of lung carcinomas with different cell proliferation and apoptosis rates \cite{mattern2004imbalance}, the spread of disruptions in railway systems \cite{johansson2011vulnerability}, the collective behaviour in online social movements \cite{adamic2015diffusion}, and so forth. In general, our model has a strong potential in characterizing a variety of complex systems with binary state components and network structure, on which non-monotone dynamics are running, and we are going to explore those candidate systems in future study.

\paragraph*{}
On the other hand, theoretically, in active phase with high ratio, after being able to calculate the average activity value given a group of parameters, the next question we want to ask naturally is how the activity value fluctuates around the average temporally. Similar to the corresponding result of Maslov sandpile model \cite{kendal2015self}, we have found that the fluctuation sizes of activity value could be properly described by Tweedie compound Poisson distribution which is a member of a family of statistical models called Tweedie family. With the knowledge of the fluctuation size distribution we can further investigate the failure of networks due to finite size effect and random fluctuation even when the ratio of recovery/death rates is above the critical ratio $R^{(c)}$, which is an important addition to the complete picture of the robustness of networks under cascade-recovery dynamics.




\section*{Acknowledgements}
The authors would like to thank Ginestra Bianconi, James Gleeson, Gunnar
Pruessner, David Steinsaltz, Yi Yang and Qing Yao for very insightful
discussion.

\newpage
\bibliographystyle{unsrt}
\bibliography{thrsh-rcv}

\end{document}